\documentclass[11pt, 3p]{elsarticle}

\usepackage{mathpazo}
\usepackage[T1]{fontenc}
\usepackage[utf8]{inputenc}
\usepackage{multirow}
\usepackage[colorlinks=true]{hyperref}
\usepackage[english]{babel}
\usepackage{adjustbox}
\usepackage{xcolor}
\usepackage{booktabs}
\usepackage{amsmath}
\usepackage{textcomp}
\usepackage{lineno}
\usepackage{graphicx}
\usepackage{tikz}
\usepackage{siunitx}
\usepackage{subcaption}
\usepackage{setspace}

\usepackage{caption}
\usepackage[font=small,labelfont=bf, skip=0pt]{caption}
\biboptions{sort&compress}

\biboptions{sort&compress}

\graphicspath{{./}} 

\begin{document}
	
\begin{frontmatter}

\journal{Scripta Materialia}

\title{Synergism between B and Nb improves fire resistance in microalloyed steels}

\author[1]{Pedro P. Ferreira}
\author[2]{Felipe M. Carvalho}
\author[3]{Edwan A. Ariza-Echeverri}
\author[3]{Pedro M. Delfino}
\author[3]{Luiz F. Bauri}
\author[3]{Andrei M. Ferreira}
\author[2]{Ana P. V. Braga}
\author[1]{Luiz T. F. Eleno\corref{cor}}
\cortext[cor]{Corresponding author:}
\ead{luizeleno@usp.br}
\author[3]{Hélio Goldenstein}
\author[3]{André P. Tschiptschin\corref{cor}}
\ead{antschip@usp.br }

\address[1]{Computational Materials Science Group (ComputEEL/MatSci), Department of Materials Engineering, Lorena School of Engineering, Universidade de S{\~a}o Paulo, DEMAR, Lorena--SP, Brazil}
\address[2]{Institute of Technological Research (IPT), São Paulo--SP, Brazil}
\address[3]{Department of Metallurgical and Materials Engineering, Escola Politécnica, Universidade de São Paulo, São Paulo--SP, Brazil}

\begin{abstract}


The development of new fire-resistant steels represents a challenge in materials science and engineering of utmost importance. Alloying elements such as Nb and Mo are generally used to improve the strength at both room- and high-temperatures due to, for example, the formation of precipitates and harder microconstituents. In this study we show alternatively that the addition of small amounts of boron in Nb-microalloyed steels may play a crucial role in maintaining the mechanical properties at high temperatures. The 66\,\% yield-strength criteria for fire resistance is achieved at $\approx 574$\,°C for a boron steel, whereas without boron this value reaches $\approx 460$\,°C, a remarkable boron-induced mechanical strengthening enhancement. DFT calculations show that boron additions can lower the vacancy formation energy when compared to pure ferrite and, for Nb-B steels, there is a further 24\,\% reduction, suggesting that the boron-niobium combination acts as an effective pinning-based strengthening agent.

\end{abstract}


\end{frontmatter}


Alloying is an essential tool to design and engineer new materials with optimized properties. For decades, boron (B), for instance, has been added to several classes of steels to improve its hardenability \cite{lin1987, tsuji1997, gao2015, hu2015, koley2018, sharma2019}. It is added to steels between 10 to 30\,ppm to optimize ultimate tensile strength and toughness. When added to steels up to this limit, boron segregates to austenite grain boundaries (AGBs), reducing the grain boundary energy and increasing the hardenability by suppressing the nucleation of allotriomorphic ferrite  \cite{miyamoto2018}. Due to the boron's high affinity with N and C, higher amounts of boron may induce boron nitrides or boron carbides at the AGBs, acting as nucleation sites for ferrite at high temperatures \cite{koley2018}. In addition, steels containing B are less susceptible to distortion and quenching cracking during heat treatments \cite{ghali2012}. Recently, Sharma et al. \cite{sharma2019_2} published a review of the effects of B on heat-treatable steels, focusing on some critical aspects. For instance, B can present equilibrium segregation, i.e., redistribution of the solute impurities when the temperature increases, causing a decrease in the amount of B solute at the grain boundaries and consequently in the hardenability, or it can present non-equilibrium segregation by the formation of vacancy-boron that migrates to the grain boundaries. Moreover, B can create a synergic interaction with some chemical species, such as Mo and Nb, and B-vacancy or complex precipitates, such as B$_2$O$_3$, BN, M$_{23}$(B,C)$_6$, and M$_7$(B,C)$_3$. Therefore, B needs to be protected during steel-making production and heat treatment processes to avoid its precipitation and consequently loss of effectiveness in hardenability. One of the most usual procedures is to add Ti or Al to protect this effectiveness. In this case, Ti forms the more stable TiN instead of BN, ensuring that boron remains in solid solution. Ali et al. \cite{ali2021} studied the potential factors that affect the B hardenability in low carbon alloyed steels, concluding that the formation of coarse precipitates of M$_{23}$(B,C)$_6$ in AGBs led to the deterioration of toughness and affected the hardenability negatively. 

Also an important ingredient for the strength of steels, the hindering of dislocation mobility is highly desirable to improve the mechanical properties at elevated temperatures since it mitigates the dislocation annihilation with increasing temperature. Recently, Jo et\,al. \cite{jo2020} proposed that the addition of transition metal solutes, especially Mo and Nb elements, significantly lowers the vacancy formation energy in Fe-BCC, paving the way for these lattice defects to be easily formed. Notably, the predicted fraction of vacancy-related defects was indeed confirmed through positron annihilation lifetime spectroscopy by the same authors \cite{jo2020}. These observations led us to search for novel solute combinations through first-principles electronic-structure calculations\footnote{First-principes spin-polarized electronic-structure calculations were performed within the Kohn-Sham scheme \cite{kohn1965} of the Density Functional Theory (DFT) \cite{hohenberg1964} with scalar-relativistic  projector augmented wave pseudopotentials \cite{dal2014} as implemented in Quantum \textsc{Espresso} \cite{giannozzi2009,giannozzi2017}. Exchange and Correlation (XC) effects were treated with the generalized gradient approximation (GGA) according to Perdew-Burke-Ernzerhof (PBE) parametrization \cite{perdew1996}. We used a kinetic energy cutoff of 70\,Ry (1\,Ry $\approx$ 13.6\,eV) for wave functions and 1120\,Ry for the charge density and potential. The Monkhorst-Pack scheme \cite{monkhorst1976} was used for a $6\times6\times6$ $k$-point sampling in the first Brillouin zone. Self-consistent-field (SCF) calculations with $10^{-6}$\,Ry convergence threshold were carried out using the Marzari-Vanderbilt smearing \cite{marzari1999} with a spreading of 0.01\,Ry for Brillouin-zone integration. A mixing factor of 0.1 with 12 iterations in charge density Broyden mixing scheme was used to reach the convergence threshold for self-consistency. All lattice parameters and internal degrees of freedom were relaxed in order to guarantee a ground-state convergence of 10$^{-4}$\,Ry in total energy, 10$^{-4}$\,Ry/a$_0$ ($a_0\approx0.529\,$\AA) for forces acting on the nuclei, and 10$^{-6}$\,Ry/a$_0^3$ for stresses acting on the crystal structure. For further details on supercell calculations, see, for instance, Refs.\,\cite{ram2012, dorini2021}.}, aiming to optimize this mechanism. We have discovered that small additions of B and Nb to a specific steel composition surprisingly leads to very significant maintenance of its mechanical properties at fire-level temperatures. Such characteristics can be used with advantage in devising new, cheaper steels with reduced or no molybdenum contents with improved high-temperature properties, leading ultimately to safer buildings and other structures in which safety is a primordial concern.
 
\begin{figure*}[t]
	\centering
	\begin{subfigure}{0.5\textwidth}
		\centering
        \includegraphics[width=.5\textwidth]{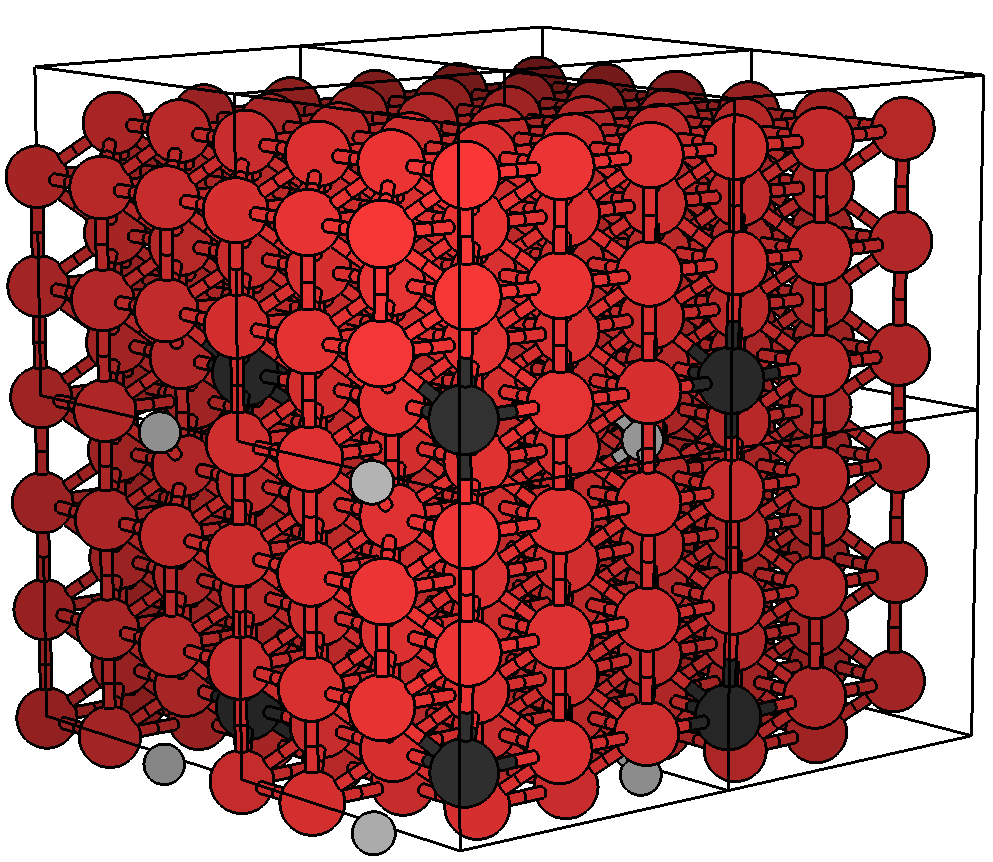}
        \includegraphics[width=.42\textwidth]{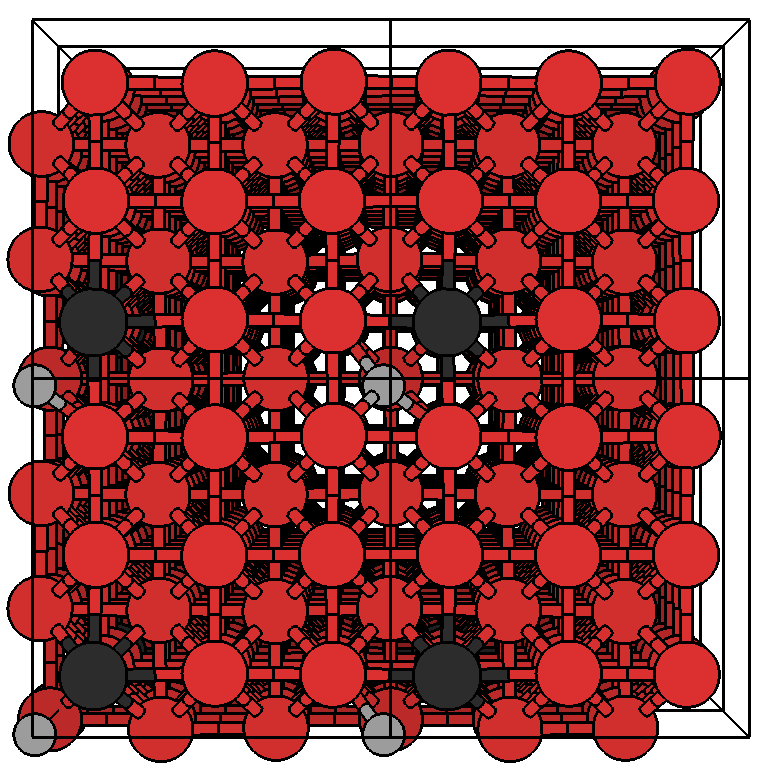}\\
        \includegraphics[width=.5\textwidth]{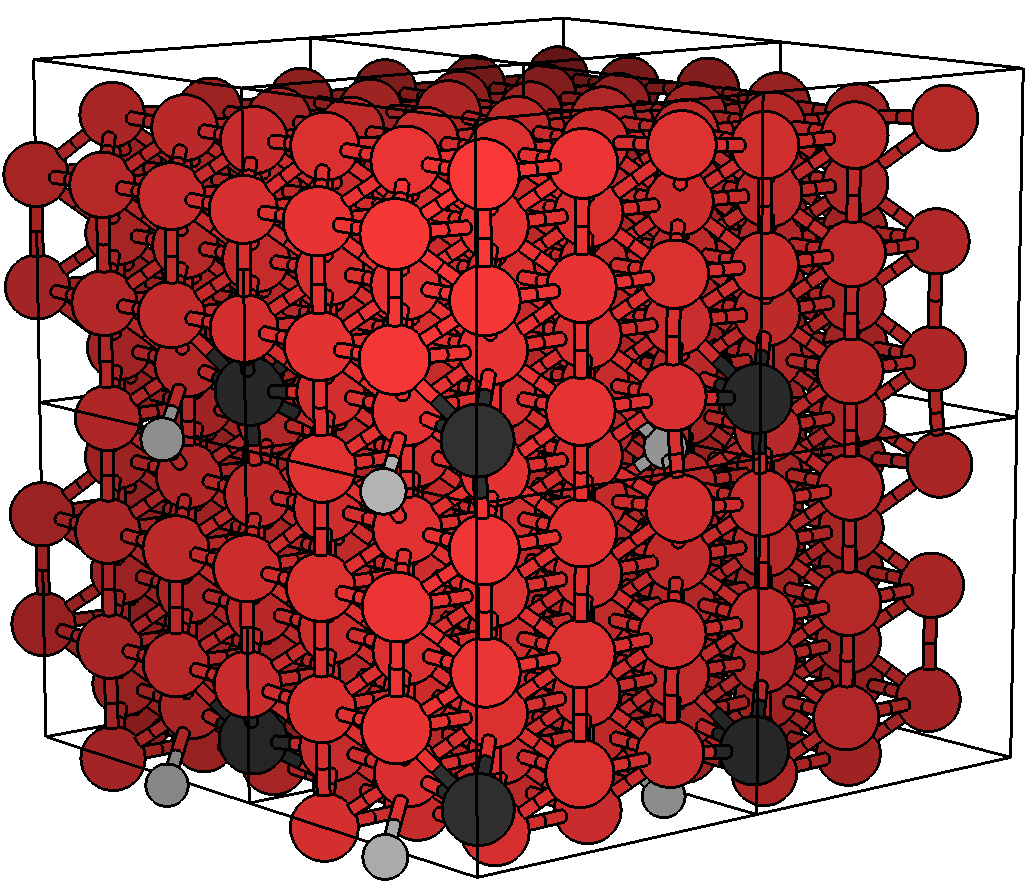}
        \includegraphics[width=.42\textwidth]{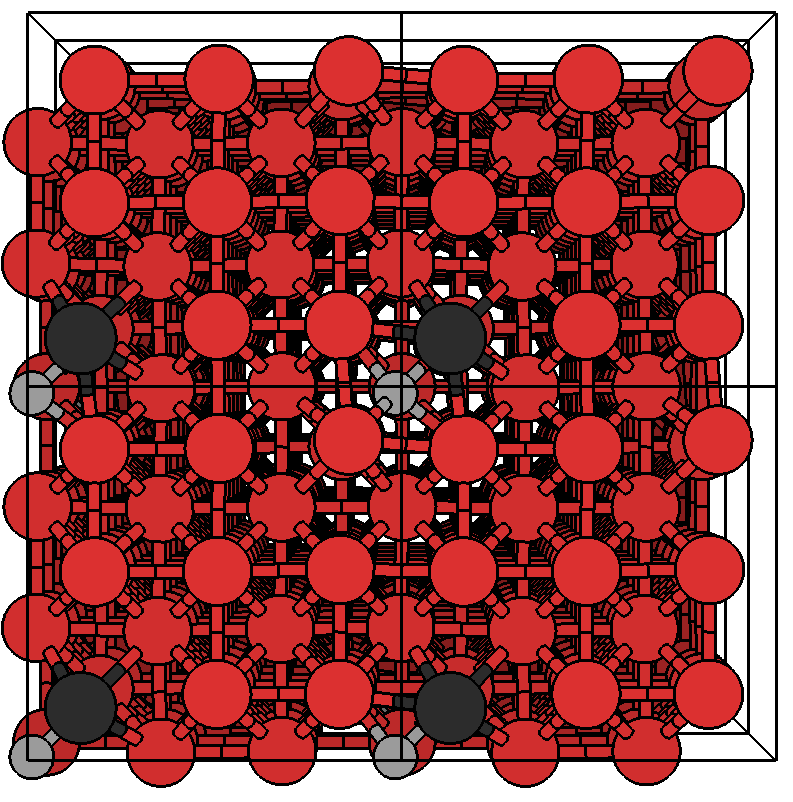}
		\caption{}
	\end{subfigure}%
	\begin{subfigure}{.5\textwidth}
		\centering
        \includegraphics[width=.99\textwidth]{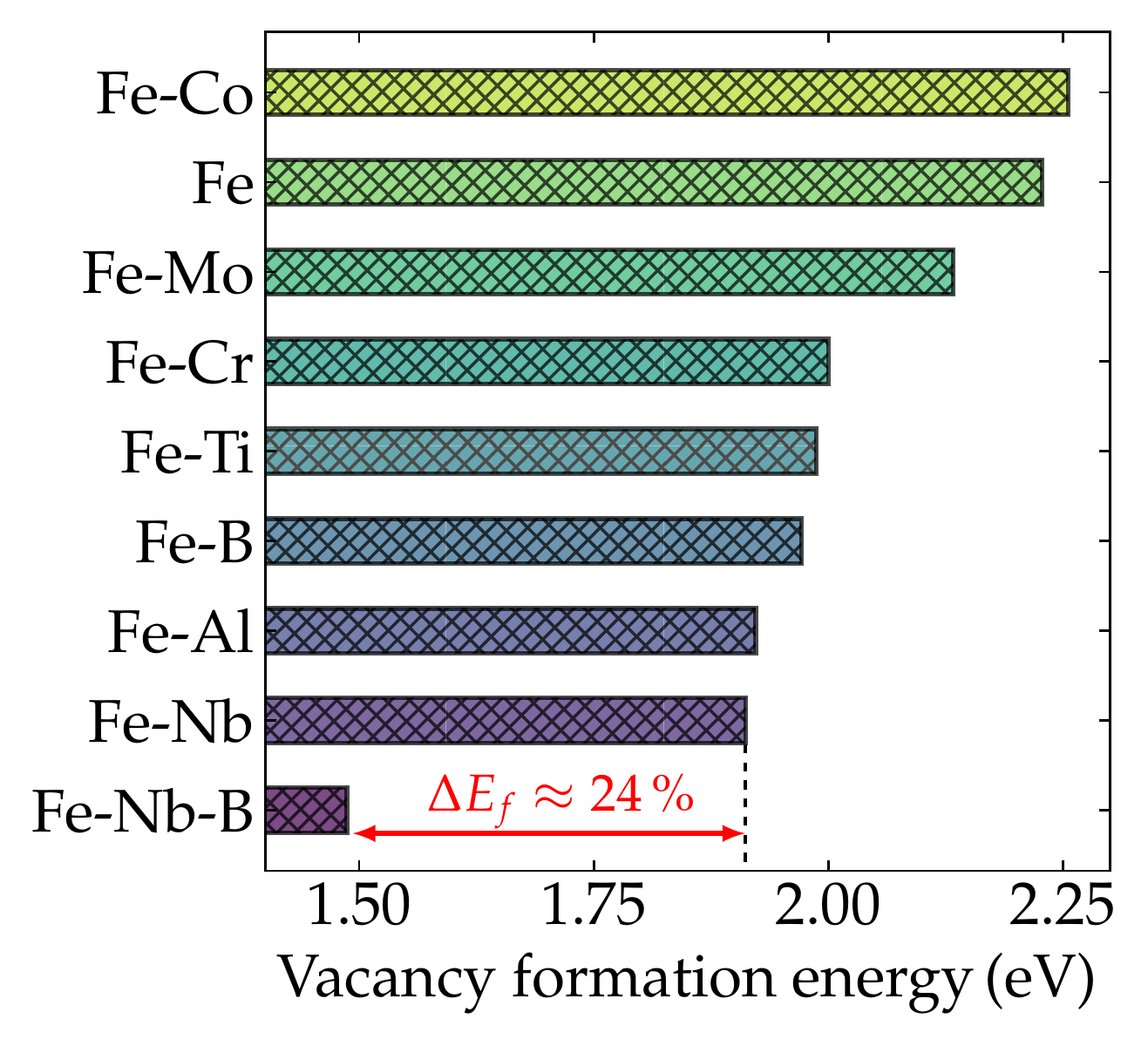}
		\caption{}
	\end{subfigure}
    \caption{First-principles calculations of the Fe-Nb-B structural system. (a) Representation of the $3\times3\times3$ supercell used in our calculations. Fe atoms are shown in red, B atoms in light gray, and other solutes in dark gray. The upper panel shows the supercell structures without vacancy, while the lower panel contains the supercell with a vacancy point defect. (b) DFT-calculated vacancy formation energy of BCC-Fe with several solutes. The difference between the energy required to form a vacancy in Fe-Nb and Fe-Nb-B is almost 25\,\%, as indicated by the red arrow in the figure.}
	\label{fig:theory} 
\end{figure*}

According to our results, summarized in Figure \ref{fig:theory}, it is expected that Nb could reduce almost 16\,\% of the vacancy formation energy compared to pure ferrite (2.23\,eV), or 11\,\% compared to those alloys containing Mo (2.13\,eV), in remarkable agreement with previous theoretical calculations \cite{jo2020}. However, we have found additionally that when boron is incorporated into Fe--Nb system, the energy required for a vacancy to be formed becomes 1.49\,eV, representing an expressive lowering as great as 24\,\% compared to the situation without boron (1.91\,eV). These results reveal that B-induced higher density of vacancy-related defects can act effectively as a trapping barrier to the dislocation mobility, and, consequently, as a pinning-based steel strengthening mechanism. For the sake of comparison, the degree of decrease of vacancy formation energy promoted by B solid solution in Nb microalloyed ferritic steels represents an enhancement of 3 orders of magnitude in the amount of vacancy point defects, assuming an Arrhenius' law behavior. Therefore, such enhancement must be imperative to the macroscopic properties of this class of steels. Furthermore, as also pointed out by Jo et\,al. \cite{jo2020}, the formation of solute-vacancy pairs and related lattice defects were also observed in other iron binary alloys with refractory metals \cite{wang2010, idczak2012, idczak2014, idczak2014_2}, further reinforcing their importance in high-temperature resistance. Our results put a new piece in the puzzle, with B being a potentially central actor in the ongoing fire-resistant steels endeavor.

\begin{figure}[t]
	\centering
        \includegraphics[width=.65\textwidth]{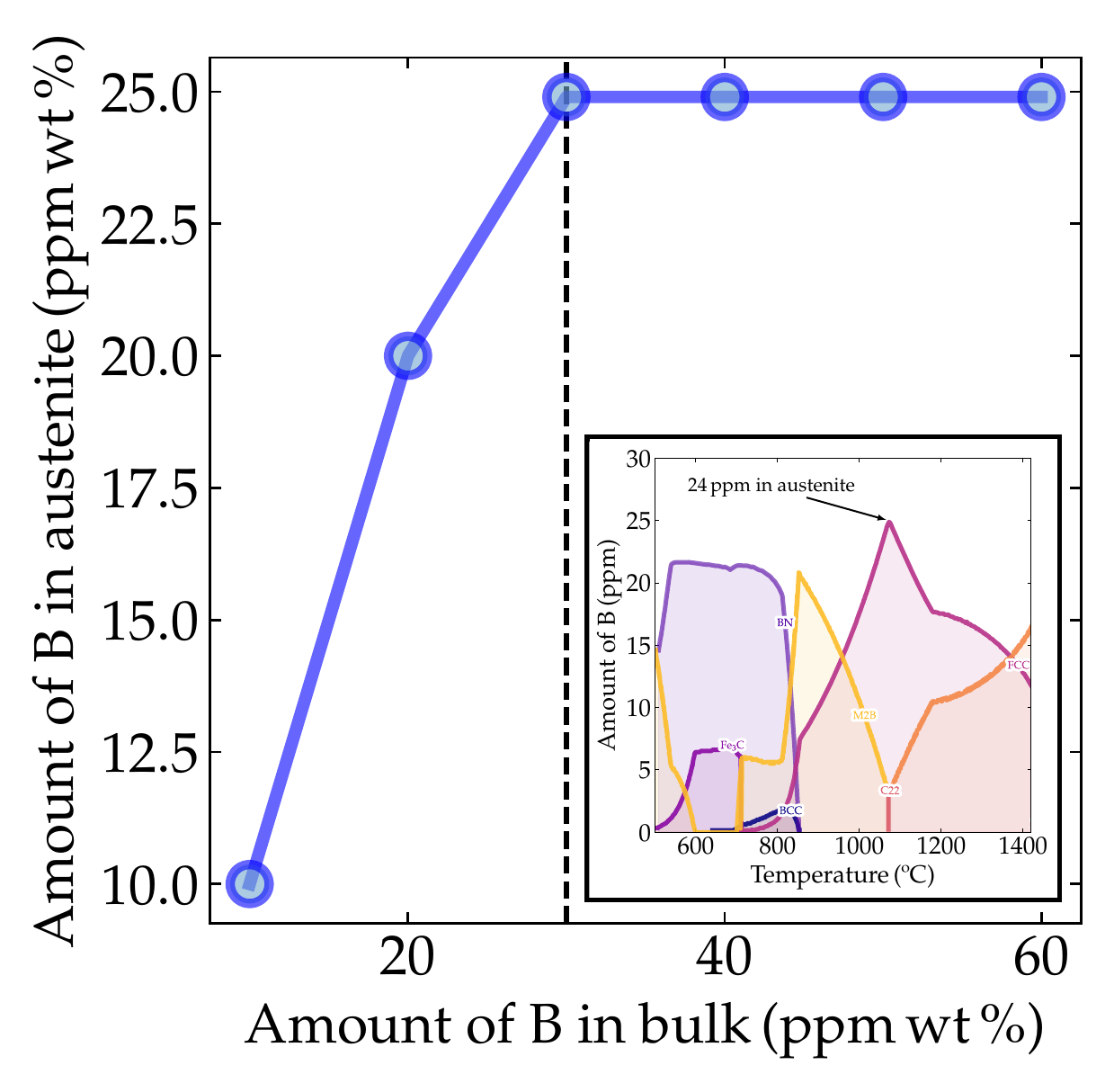}
    \caption{Thermodynamic prediction of the amount of boron in austenite as a function of bulk composition. The inset depicts the thermodynamic simulation of the boron content in different microconstituents as a function of temperature, showing a broad peak around 24.9\,ppm boron solubility in austenite for the bulk (Nb,B)-microalloyed steel.}
	\label{fig:calphad} 
\end{figure} 

However, for reaching the above predicted behavior it is important to find B predominantly as solid solution. Therefore, in order to probe in which phase boron atoms are predicted to be solubilized, we have conducted thermodynamic simulations using Thermocalc® (TCFE11-21b database). The simulations were conducted using the base chemical composition, given in Table \ref{tab:composition}, but varying the boron contents as 10, 20, 28 (the \emph{a priori} experimentally desired composition), 30, 40, 50 and 60\,ppm. Figure \ref{fig:calphad} shows the temperature-dependent maximum amounts of boron within the phases for the boron-added alloys. One can see that it is possible to solubilize 24.9\,ppm of B in the austenite from a global of 28\,ppm. Running the same simulation for various boron contents we have verified how the solubility of boron in austenite behaves as a function of the amount of boron added to the material. Notably, the boron solubility limit in austenite is 24.9\,ppm regardless of the amount of boron added to the alloy composition.

All these predictions thus motivate us to further investigate experimentally the simultaneous effect of B and Nb on improving the mechanical properties at high temperatures. Ingots of 80\,kg of two alloys were prepared in a vacuum induction furnace (VIF) to produce the samples. The chemical composition of the materials studied is presented in Table \ref{tab:composition}, one with Fe-C-Mn-Nb-B (with B) and the other with Fe-C-Mn-Nb (without B).

\begin{table*}
	\centering
	\caption{Chemical composition (in wt.\,\%) of the two ingots used for the experimental investigation in the present work.}
	\label{tab:composition}
\begin{tabular}{lcccccc}
\toprule
Alloy & B & C & Nb & Mn & N & Ti \\
\midrule
 with B & 0.0028 & 0.05 & 0.10 & 1.00 & 0.0029 & <0.015 \\
without B & 0.00 & 0.05 & 0.10 & 1.00 & 0.0029 & <0.015 \\
\bottomrule
\end{tabular}
\end{table*}

Firstly, a plate from each ingot was machined, and then it was subjected to a controlled thermomechanical rolling treatment. The reheating temperature was 1250\,°C, which promotes the dissolution of niobium carbides allowing precipitation during the hot-rolling treatment. The thermomechanical process was performed with six rolling steps of 76\,\% total reduction thickness, ending above the Ac$_1$ temperature. After hot rolling, the plates were air-cooled to room temperature. Then, isothermal tensile tests were carried out at room temperature (RT) and 400, 500, 600, 700, and 800\,°C, to evaluate the yield strength (YS) at 0.2\,\% offset at those different temperatures. Mechanical test specimens were taken from the plates in the rolling direction according to ASTM E8/E8M \cite{ASTME8} for the RT test and ASTM E21 \cite{astme21} for high temperature (HT) tests. For the isothermal high-temperature tensile tests, the samples were heated to the desired temperature under a heating rate of 10\,°C/min, maintained for 15 minutes for thermal stabilization, and then a strain rate of 0.00416\,s$^{-1}$ was applied until fracture. Finally, after fracture, the samples were air-cooled to room temperature.  

\begin{figure*}[h]
	\centering
    \includegraphics[width=.49\textwidth]{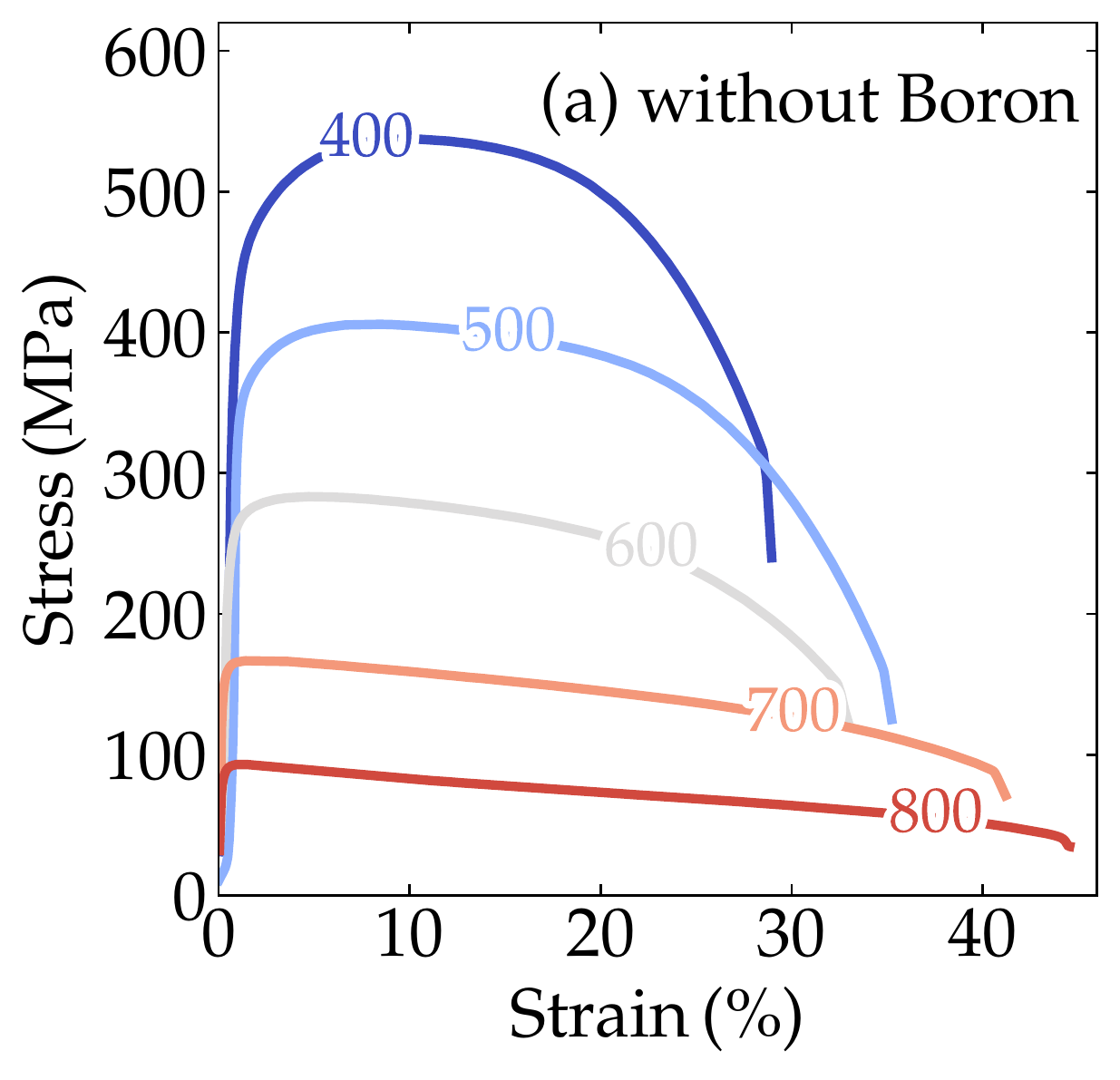}
    \includegraphics[width=.49\textwidth]{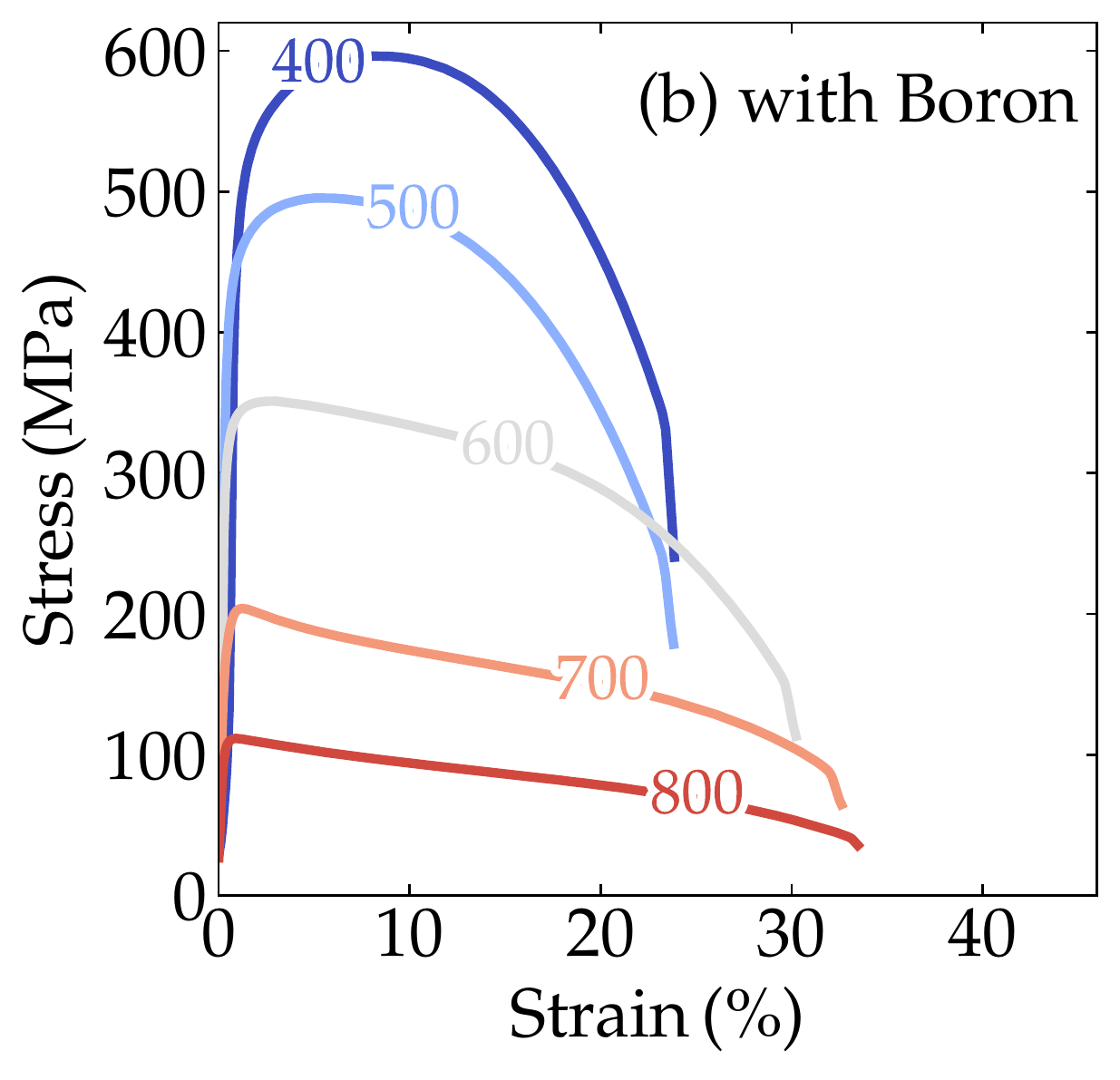}
    \caption{The high-temperature static tests for the alloy (a) without boron and (b) with added boron.}
	\label{fig:tensile-test} 
\end{figure*}

Figures \ref{fig:tensile-test}(a) and \ref{fig:tensile-test}(b) show the tensile test results of the boron-containing and non-containing alloys, performed at 400, 500, 600, 700 and 800\,°C. Both alloys show a reduction in the yield stress associated with increased ductility with increasing temperature. On the other hand, for the alloy with boron, a less severe decay of the yield stress with increasing temperature occurs.

In addition to the isothermal high-temperature tensile tests, heating with a constant load applied, also known as accelerated creep tests, were performed, following the procedure delineated in Walp’s investigation \cite{walp2003}, which consists of using first a constant load equivalent to a fraction of yield strength at room temperature and then heating the system at a set rate. While the system heats up, the material yield stress decreases, and when a specific imposed stress (constant load) exceeds the material yield stress, the material plastically deforms. As a result, a strain curve can be obtained as a function of temperature. A failure criterion of 1\,\% strain was adopted in the present study, which is consistent with previous reports \cite{speer2015}. A standard creep machine was used to perform the heating, adopting a constant load of 188\,MPa (the equivalent of $ \approx\,35\%$ YS$_{RT}$ of the alloys) and a heating rate of 5\,°C/min. 

\begin{figure*}[t]
	\centering
	\begin{subfigure}{0.49\textwidth}
    \includegraphics[width=\textwidth]{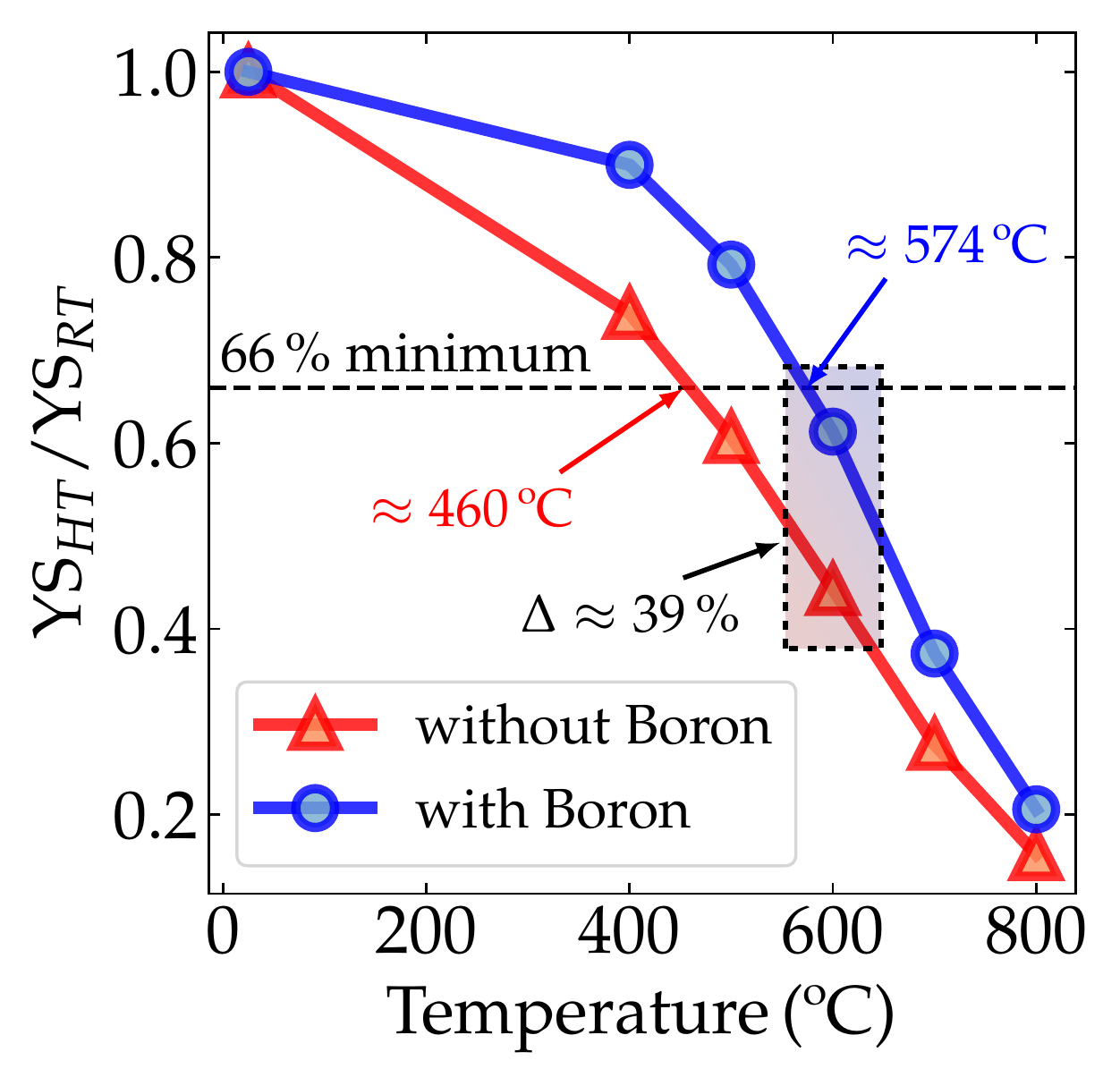}
    \caption{}
    \end{subfigure}
   	\begin{subfigure}{0.49\textwidth}
    \includegraphics[width=\textwidth]{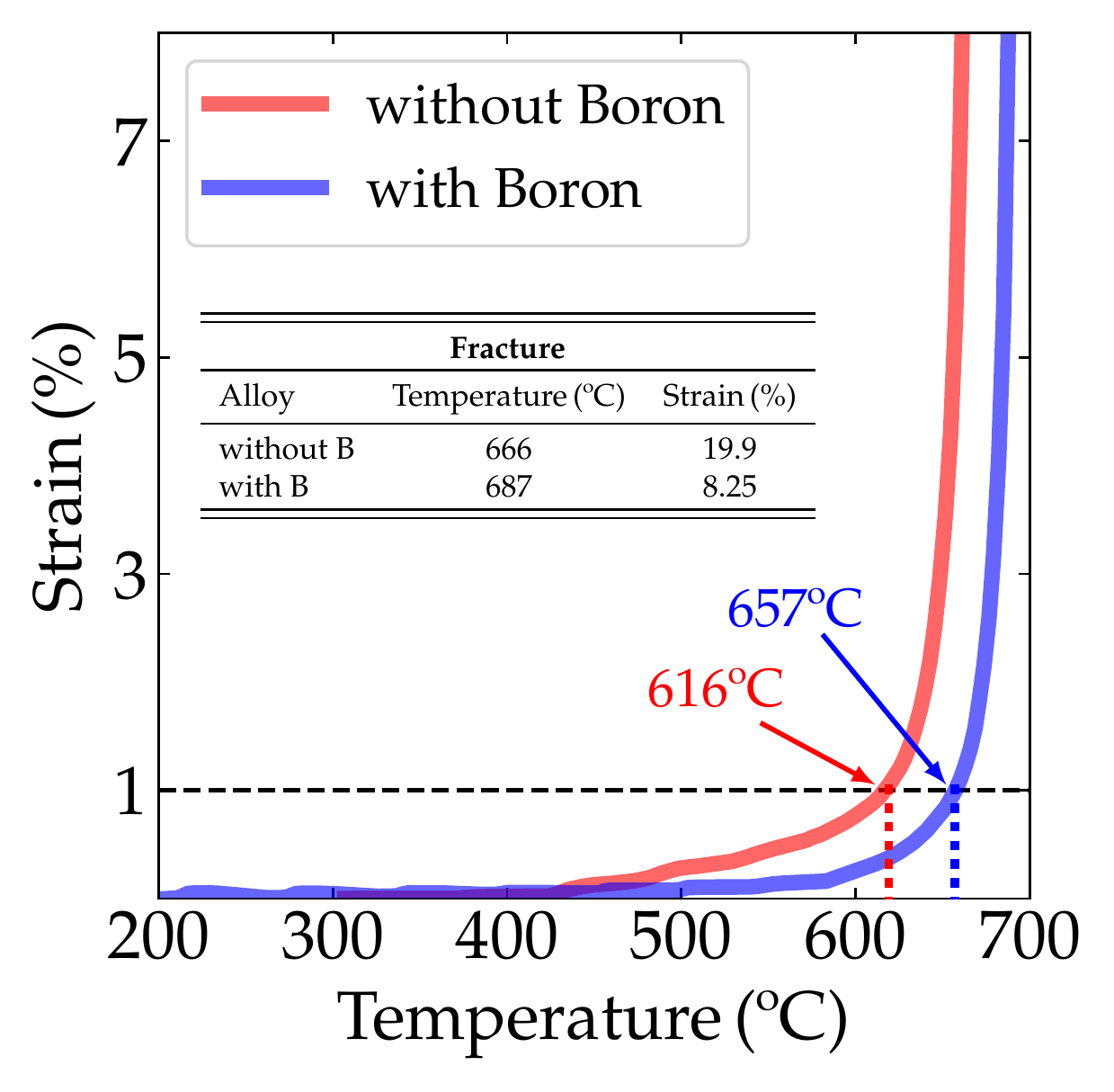}
    \caption{}
	\end{subfigure}
    \caption{(a) Yield strength ratio of the alloys without boron (alloy 1) and with boron-added (alloy 2) at room temperature (RT or 25\,°C), 400, 500, 600, 700 and 800\,°C; minimum threshold indication of 66\,\% for yield strength ratio at 600\,°C with their respective values. (b) Plastic strain with temperature from transient test data -- heating rate: 5\,°C/min, preload applied: 188\,MPa (equivalent to 33\,\% YSRT­ for alloy 1-without B and 36\,\% YSRT for alloy 2-with B).}
	\label{fig:strenght-ratio} 
\end{figure*}

The variation on yield stress with temperature is shown in Figure \ref{fig:strenght-ratio}(a). The literature on fire-resistant structural steels presents several methods to measure the fire resistance of a given material. Based on the ASTM A1077/A1077M standard \cite{astma1077}, the literature \cite{walp2003, chijiiwa1993} sets a criterion to determine whether structural steel is fire resistant or not, based on the ratio between the yield strength measured at 600 °C (YS$_{HT}$) and the yield strength measured at room temperature (YS$_{RT}$). For a steel to be considered fire-resistant, this ratio (YS$_{HT}$/YS$_{RT}$) must be greater than 2/3, or 66\,\%. In Figure \ref{fig:strenght-ratio}(a), one can see the variation of the ratio as a function of temperature for the two alloys studied.

From the graphic shown in Figure \ref{fig:strenght-ratio}(a), it is possible to verify that the boron-added alloy presents a lower yield strength at room temperature than its counterpart without boron. However, the boron-added steel maintains a higher yield strength ratio with increasing temperature. Indeed, in the detached rectangle in Figure \ref{fig:strenght-ratio}(a) it is clear that YS$_{HT}$/YS$_{RT}$ for the boron-added alloy is significantly higher than for those without boron. 

The strain vs. temperature data obtained from the constant load test is shown in Figure \ref{fig:strenght-ratio}(b), where the alloy with boron exhibits a better performance at fire-level temperatures. This experiment simulates the fire condition on which a structural beam is subjected to constant load while the temperature increases. In this experiment, the fire resistance may be understood as the temperature required to achieve the 1\,\% strain criterion. Above 450\,°C, considering a fixed strain, the alloy with boron shows higher temperature resistance than the alloy without boron. For instance, considering the 1\,\% strain criterion as proposed by Speer et al. \cite{speer2015}, the B-added alloy resists up to 657\,°C while its counterpart without boron resists up to 616\,°C. Therefore, our experiments show that Nb-microalloyed steels containing small amounts of boron presents a better fire resistance than those without boron in their composition considering the 1\,\% strain criterion. Considering the fracture point, the alloy with boron will fracture at higher temperatures when compared to the alloy without boron. That may be associated with hardening mechanisms still acting at high temperatures in this material, such as stability of precipitates \cite{wan2012, jo2020, lee2000, lee2002} and a solid solution of B, enabling the alloy to maintain a considerable part of its yield strength at those temperatures.

To summarize, our work shows that simultaneous additions of boron and niobium to microalloyed steels can lead to significant better fire-resistance performance from both constant load test and isothermal static tests, compared to their non-boron counterparts. The 66\,\% YS ratio criterion at RT is achieved at $\approx 574$\,°C for the boron-added Nb-microalloyed steel studied, bridging the way to achieving fire-resistant steel compositions with less or even no molybdenum. As shown by DFT calculations, the addition of boron to a ferritic matrix significantly improves the stabilization of point defects by lowering the vacancy formation energy in the presence of niobium, which may play an essential role in reducing dislocation mobility. In this way, we expect that our work motivates further investigations regarding the role of boron and niobium in the development of fire-resistant steels.
    
\section*{Acknowledgments}
    
We gratefully acknowledge the financial support of the São Paulo Research Foundation (FAPESP) under Grants 2019/05005-7 and 2020/08258-0. The research was partially carried out using high-performance computing resources made available by the Superintendência de Tecnologia da Informação (STI), Universidade de São Paulo. This study was financed in part by the Coordenação de Aperfeiçoamento de Pessoal de Nível Superior -- Brasil (CAPES) --- Finance Code 001 and with the support of CNPq (National Council for Scientific and Technological Development -- Brazil). The authors also acknowledge the National Laboratory for Scientific Computing (LNCC/MCTI, Brazil) for providing HPC resources of the SDumont supercomputer, which have contributed to the research results reported in this manuscript. This research work has been supported by Companhia Brasileira de Metalurgia e Mineração (CBMM) and Empresa Brasileira de Pesquisa e Inovação Industrial – \mbox{EMBRAPII} MCE Contract 19.5.02.


\begin{thebibliography}{34}
\providecommand{\natexlab}[1]{#1}

\bibitem[{Lin and Cheng(1987)}]{lin1987}
H.-R. Lin, G.-H. Cheng.
\newblock Mater. Sci. Technol. 3 (1987) 855--859.

\bibitem[{TSUJI et~al.(1997)TSUJI, MATSUBARA, SAKAI, and SAITO}]{tsuji1997}
N.~TSUJI, Y.~MATSUBARA, T.~SAKAI, Y.~SAITO.
\newblock ISIJ Int. 37 (1997) 797--806.

\bibitem[{Gao et~al.(2015)Gao, Xue, and Yang}]{gao2015}
Y.-l. Gao, X.-x. Xue, H.~Yang.
\newblock Crystals 5 (2015) 592--607.

\bibitem[{Hu et~al.(2015)Hu, Du, Ma, Sun, Xie, and Misra}]{hu2015}
J.~Hu, L.-X. Du, Y.-N. Ma, G.-S. Sun, H.~Xie, R.~Misra.
\newblock Mater. Sci. Eng., A 640 (2015) 259--266.

\bibitem[{Koley et~al.(2018)Koley, Karani, Chatterjee, and Shome}]{koley2018}
S.~Koley, A.~Karani, S.~Chatterjee, M.~Shome.
\newblock J. Mater. Eng. Perform. 27 (2018) 3449--3459.

\bibitem[{Sharma et~al.(2019{\natexlab{a}})Sharma, Ortlepp, and
  Bleck}]{sharma2019}
M.~Sharma, I.~Ortlepp, W.~Bleck.
\newblock Steel Res. Int. 90 (2019{\natexlab{a}}) 1900133.

\bibitem[{Miyamoto et~al.(2018)Miyamoto, Goto, Takayama, and
  Furuhara}]{miyamoto2018}
G.~Miyamoto, A.~Goto, N.~Takayama, T.~Furuhara.
\newblock Scr. Mater. 154 (2018) 168--171.

\bibitem[{Ghali et~al.(2012)Ghali, El-Faramawy, Eissa et~al.}]{ghali2012}
S.~N. Ghali, H.~S. El-Faramawy, M.~M. Eissa, et~al.
\newblock JMMCE 11 (2012) 995--999.

\bibitem[{Sharma et~al.(2019{\natexlab{b}})Sharma, Ortlepp, and
  Bleck}]{sharma2019_2}
M.~Sharma, I.~Ortlepp, W.~Bleck.
\newblock Steel Res. Int. 90 (2019{\natexlab{b}}) 1900133.

\bibitem[{Ali et~al.(2021)Ali, Nyo, Kaijalainen et~al.}]{ali2021}
M.~Ali, T.~Nyo, A.~Kaijalainen, et~al.
\newblock Mater. Sci. Eng., A 819 (2021) 141453.

\bibitem[{Jo et~al.(2020)Jo, Shin, Moon et~al.}]{jo2020}
H.-H. Jo, C.~Shin, J.~Moon, et~al.
\newblock Mater. Des. 194 (2020) 108882.

\bibitem[{Kohn and Sham(1965)}]{kohn1965}
W.~Kohn, L.~J. Sham.
\newblock Phys. Rev. 140 (1965) A1133.

\bibitem[{Hohenberg and Kohn(1964)}]{hohenberg1964}
P.~Hohenberg, W.~Kohn.
\newblock Phys. Rev. 136 (1964) B864.

\bibitem[{Dal~Corso(2014)}]{dal2014}
A.~Dal~Corso.
\newblock Comput. Mater. Sci. 95 (2014) 337--350.

\bibitem[{Giannozzi et~al.(2009)Giannozzi, Baroni, Bonini
  et~al.}]{giannozzi2009}
P.~Giannozzi, S.~Baroni, N.~Bonini, et~al.
\newblock J. Phys.: Condens. Matter 21 (2009) 395502.

\bibitem[{Giannozzi et~al.(2017)Giannozzi, Andreussi, Brumme
  et~al.}]{giannozzi2017}
P.~Giannozzi, O.~Andreussi, T.~Brumme, et~al.
\newblock J. Phys.: Condens. Matter 29 (2017) 465901.

\bibitem[{Perdew et~al.(1996)Perdew, Burke, and Ernzerhof}]{perdew1996}
J.~P. Perdew, K.~Burke, M.~Ernzerhof.
\newblock Phys. Rev. Lett. 77 (1996) 3865.

\bibitem[{Monkhorst and Pack(1976)}]{monkhorst1976}
H.~J. Monkhorst, J.~D. Pack.
\newblock Phys. Rev. B 13 (1976) 5188.

\bibitem[{Marzari et~al.(1999)Marzari, Vanderbilt, {De Vita}, and
  Payne}]{marzari1999}
N.~Marzari, D.~Vanderbilt, A.~{De Vita}, M.~C. Payne.
\newblock Phys. Rev. Lett. 82 (1999) 3296.

\bibitem[{Ram et~al.(2012)Ram, Kanchana, Vaitheeswaran, Svane, Dugdale, and
  Christensen}]{ram2012}
S.~Ram, V.~Kanchana, G.~Vaitheeswaran, A.~Svane, S.~Dugdale, N.~Christensen.
\newblock Phys. Rev. B 85 (2012) 174531.

\bibitem[{Dorini et~al.(2021)Dorini, de~Freitas, Ferreira et~al.}]{dorini2021}
T.~T. Dorini, B.~X. de~Freitas, P.~P. Ferreira, et~al.
\newblock Scr. Mater. 199 (2021) 113854.

\bibitem[{Wang et~al.(2010)Wang, Xu, Guo, Wang, and Rong}]{wang2010}
X.~Wang, W.~Xu, Z.~Guo, L.~Wang, Y.~Rong.
\newblock Mater. Sci. Eng., A 527 (2010) 3373--3378.

\bibitem[{Idczak et~al.(2012)Idczak, Konieczny, and Chojcan}]{idczak2012}
R.~Idczak, R.~Konieczny, J.~Chojcan.
\newblock Solid State Commun. 152 (2012) 1924--1928.

\bibitem[{Idczak and Konieczny(2014)}]{idczak2014}
R.~Idczak, R.~Konieczny.
\newblock Appl. Phys. A 117 (2014) 1785--1789.

\bibitem[{Idczak et~al.(2014)Idczak, Konieczny, and Chojcan}]{idczak2014_2}
R.~Idczak, R.~Konieczny, J.~Chojcan.
\newblock J. Appl. Phys. 115 (2014) 103513.

\bibitem[{{ASTM E8/E8M - 16a}(2016)}]{ASTME8}
{ASTM E8/E8M - 16a}.
\newblock ASTM International, West Conshohocken, PA  (2016) 1--30.

\bibitem[{{ASTM E21 - 17}(2017)}]{astme21}
{ASTM E21 - 17}.
\newblock ASTM International, West Conshohocken, PA  (2017) 1--24.

\bibitem[{Walp(2003)}]{walp2003}
M.~Walp.
\newblock Ph.D. thesis, Colorado School of Mines (2003).

\bibitem[{Speer et~al.(2015)Speer, Matlock, and Jansto}]{speer2015}
J.~G. Speer, D.~K. Matlock, S.~G. Jansto.
\newblock Proceedings of the Value-Added Niobium Microalloyed Construction
  Steels Symposium CBMM and TMS  (2015) 133--154.

\bibitem[{{ASTM A1077/A1077M - 14}(2014)}]{astma1077}
{ASTM A1077/A1077M - 14}.
\newblock ASTM International, West Conshohocken, PA  (2014) 1--3.

\bibitem[{Chijiiwa et~al.(1993)Chijiiwa, Tamehiro, Yoshida, Funato, Uemori, and
  Horii}]{chijiiwa1993}
R.~Chijiiwa, H.~Tamehiro, Y.~Yoshida, K.~Funato, R.~Uemori, Y.~Horii.
\newblock Nippon Steel Tech. Rep. 58 (1993) 47--55.

\bibitem[{Wan et~al.(2012)Wan, Sun, Zhang, and Shan}]{wan2012}
R.~Wan, F.~Sun, L.~Zhang, A.~Shan.
\newblock Mater. Des. 36 (2012) 227--232.

\bibitem[{Lee et~al.(2000)Lee, Hong, Park, Kim, and Park}]{lee2000}
W.~Lee, S.~Hong, C.~Park, K.~Kim, S.~Park.
\newblock Scr. Mater. 43 (2000) 319--324.

\bibitem[{Lee et~al.(2002)Lee, Hong, Park, and Park}]{lee2002}
W.-B. Lee, S.-G. Hong, C.-G. Park, S.-H. Park.
\newblock Metall. Mater. Trans. A 33 (2002) 1689.

\end{thebibliography}


\end{document}